\documentclass[pra, showpacs, twocolumn, floatfix, doi=false]{revtex4-1}

\usepackage[utf8]{inputenc}
\usepackage{graphicx}
\usepackage{amsmath, amsfonts, amssymb, bm}
\usepackage{slashed}

\renewcommand{\vec}[1]{\boldsymbol{#1}}
\newcommand{\bra}[1]{\langle {#1} |}
\newcommand{\ket}[1]{| {#1} \rangle}
\newcommand{\braket}[1]{\langle {#1} \rangle}
\newcommand{\Texp}{\mathcal{T}\exp}

\begin{document}

\title{Spin dynamics in Kapitza-Dirac scattering of electrons from bichromatic laser fields}

\author{Matthias M. Dellweg}
\author{Hani M. Awwad}
\author{Carsten M\"uller}
\affiliation{Institut f\"ur Theoretische Physik I, Heinrich-Heine-Universit\"at D\"usseldorf, Universit\"atsstr. 1, 40225 D\"usseldorf, Germany}

\date{\today}

\begin{abstract}
Kapitza-Dirac scattering of nonrelativistic electrons from counterpropagating bichromatic laser waves of linear polarization is studied.
The focus lies on the electronic spin dynamics in the Bragg regime when the laser fields possess a frequency ratio of two.
To this end, the time-dependent Pauli equation is solved numerically, both in coordinate space and momentum space.
Our numerical results are corroborated by analytical derivations.
We demonstrate that, for certain incident electron momenta, the scattering crucially relies on the electron spin which undergoes characteristic Rabi-like oscillations.
A parameter regime is identified where the Rabi oscillations reach maximum amplitude.
We also briefly discuss spin-dependent Kapitza-Dirac scattering of protons.
\end{abstract}

\maketitle

\section{Introduction}
\label{sec:intro}
The interaction between electrons and the electromagnetic fields of a laser wave mainly relies on the electron charge.
In addition, the electronic spin degree of freedom can couple to the laser field via the electron's magnetic moment.
Accordingly, when a free electron is exposed to a plane-wave laser pulse,
the electronic spin vector performs a precessive motion.
However, after the laser pulse has passed,
the electron will end up in the same spin state as it was before the interaction started \cite{Walser2002}.

Nontransient electron spin transitions can occur only if the electron undergoes a scattering process inside the laser field.
Then, the electronic spin state before and after the interaction with the field may differ.
Corresponding spin transitions have been studied theoretically, e.g.\@,
in multiphoton Compton scattering \cite{Ivanov2004,Boca2012,Krajewska2013,*Krajewska2014},
strong-field photoionization \cite{Faisal2004,Klaiber2014} and laser-assisted Mott scattering \cite{Szymanowski1998,Panek2002a}.
Typically, spin effects can become sizeable at very high field frequencies or very high field strengths.

Spin effects may also arise in Kapitza-Dirac scattering \cite{Rosenberg2004,Ahrens2012,Ahrens2013,Mcgregor2015,Erhard2015,*[{see also }]Bauke2014b,*Bauke2014a}.
The Kapitza-Dirac effect as originally proposed,
denotes the quantum mechanical diffraction of an electron beam on the periodic potential generated by a standing wave of light \cite{Kapitza1933,Fedorov1981,Batelaan2007}.
The latter can be formed by two counterpropagating laser beams.
In its original version \cite{Kapitza1933} the effect can be understood as a combined absorption and emission process involving two photons:
The incident electron absorbs one photon of momentum $\vec k$ from one of the laser beams,
and emits another photon of momentum $-{\vec k}$ into the counterpropagating beam (stimulated Compton scattering),
resulting in a momentum transfer of $2{\vec k}$.
An experimental realization of Kapitza-Dirac diffraction in its original form was accomplished some years ago,
utilizing optical laser intensities of the order of $\sim 10^{9}$--$10^{11}$\,W/cm$^2$ \cite{Freimund2001,*Freimund2002}.
Related experiments observed the Kapitza-Dirac effect on atoms \cite{Bucksbaum1988,Gould1986,Martin1988}.

Distinct spin effects in Kapitza-Dirac scattering were predicted when a moderately relativistic
electron beam impinges under the (generalized) Bragg angle on a standing x-ray laser wave \cite{Ahrens2012}.
In contrast to the original work \cite{Kapitza1933}, this spin-dependent Kapitza-Dirac effect relies on a 3-photon process,
where two photons are absorbed and one photon is emitted (or vice versa).
The interaction may be considered as arising from an $e^2{\vec A}^2$ term in the Hamiltonian,
in combination with a ${\vec \sigma}\cdot{\vec B}$ term.
Hence, the coupling involves both the electric charge and the magnetic moment of the electron.
While the theory in \cite{Ahrens2012,Ahrens2013} was based on the Dirac equation,
a relativistic treatment of the Kapitza-Dirac effect based on the Klein-Gordon equation may be found in \cite{Haroutunian1975,*Fedorov1980}.

Electron-laser interaction dynamics can be enriched further in bichromatic fields containing two different frequency components \cite{Ehlotzky2001}.
In particular, characteristic quantum interferences may arise when the frequencies are commensurate.
Two-color effects have been studied for a variety of processes, comprising strong-field photoionization \cite{Schafer1992,*Yin1992,*Schumacher1994,*Veniard1995,*Paulus1995},
laser-assisted electron-atom scattering \cite{Varro1993,*Kaminski1995,*Milosevic1997},
and x-ray Thomson scattering \cite{Sperling2013}.
Quantum interference and relative phase effects have recently also been revealed for Kapitza-Dirac scattering in bichromatic and multichromatic standing laser waves \cite{Dellweg2015,Vidil2015}.

Bichromatic laser fields may be exploited to facilitate spin effects in Kapitza-Dirac scattering.
Recently it has been demonstrated \cite{Mcgregor2015} that spin flips in 3-photon processes similar to
\cite{Ahrens2012,Ahrens2013} can arise in the nonrelativistic regime of low electron energies and field frequencies
when the electron scatters from two counterpropagating laser beams which possess a frequency ratio of two (see also \cite{Freimund2003}).
The spin effects are most pronounced when the incident electron momentum is perpendicular to the linear field polarization.
Otherwise, 3-photon scattering involving the ${\vec p}\cdot{\vec A}$ interaction term would generally dominate \cite{Smirnova2004}.
The occurence of coherent electron scattering in this field configuration
may be understood by noting that the system can be Lorentz transformed into a frame of reference
where the counterpropagating waves possess equal frequencies and form a standing wave.

We point out that spin effects in the original (2-photon) Kapitza-Dirac effect were found to be very small \cite{Rosenberg2004}, but can become distinct when elliptically polarized fields are applied \cite{Erhard2015}.
In this case, however, the spin-flip transitions compete with spin-preserving electron scattering and are suppressed by a relative factor of $\hbar k/mc$.

In this paper, we study spin-dependent Kapitza-Dirac scattering of nonrelativistic electrons from counterpropagating bichromatic laser fields of linear polarization.
The electron spin dynamics is revealed by numerically solving the Pauli equation, both in coordinate space and in momentum space.
Our main goal is to demonstrate the time evolution of the spin-dependent scattering probabilities.
This way, the recent results in \cite{Mcgregor2015} are extended.
We show that the electron spin undergoes characteristic Rabi-like oscillations and derive an analytical formula for the
corresponding Rabi frequency within the framework of time-dependent perturbation theory.
A parameter regime is identified where the Rabi oscillations reach their full amplitude.
In addition, we briefly discuss spin-dependent Kapitza-Dirac scattering of protons and highlight the relevance of the particle's $g$-factor for the process.

It is worth mentioning that analogies of Kapitza-Dirac scattering also exist in other systems:
for example, electron scattering from travelling waves in dielectric media \cite{Avetissian1976,*Hayrapetyan2015} or
coherent electron scattering by optical near-fields in transmission microscopes \cite{Feist2015}.
Besides, the application of counterpropagating bichromatic laser waves with specific frequency difference has been proposed as an interferometric beam splitter \cite{Marzlin2013}.

Our paper is organized as follows.
In Sec.~\ref{sec:theory} we present our theoretical formalism based on the Pauli equation.
Also a perturbative treatment of the spin-dependent three-photon Kapitza-Dirac effect is provided in Sec.~\ref{sub:dyson},
resulting in an analytical formula for the spin-flip Rabi frequency.
In Sec.~\ref{sub:effective_potential} and App.~\ref{sec:volkov},
the latter result is confirmed by an alternative treatment which relies on a Magnus expansion of the time-dependent Pauli equation,
and a third approach involving relativistic Volkov states of the Dirac equation.
Our numerical results are presented in Sec.~\ref{sec:numerical}.
First, we show the spin-resolved time evolution of an electron wave packet in the counterpropagating bichromatic laser waves.
Then we present Rabi oscillation dynamics for various field parameters using plane-wave electrons.
We generalize our results with a parameter scan to identify regions of different behavior.
In Sec.~\ref{sec:proton}, a comparison between spin-dependent Kapitza-Dirac scattering of electrons versus protons is drawn.
We finish with concluding remarks in Sec.~\ref{sec:conclusion}.

\section{Theoretical framework}
\label{sec:theory}
\subsection{Basic equations}
\label{sub:basic_eqn}
The nonrelativistic domain of electron-light scattering, including the spin degree of freedom, is described by the time-dependent Pauli equation
\begin{equation}
 i \frac{\partial}{\partial t} \psi = \left[ \frac{1}{2 m} \left( -i\vec\nabla + \frac{e}{c} \vec{A} \right)^2 + \mu_{\rm B}\, \vec{\sigma}\cdot\vec{B}\right] \psi
\label{eqn:tdpe}
\end{equation}
where $m$ is the electron mass, $-e$ its charge, $c$ the speed of light, $\mu_{\rm B} = \frac{e}{2mc}$ the Bohr magneton, and $\psi$ the electron wave function.
We have set $\hbar$, the reduced Planck constant to unity for convenience.
Besides, $\vec A$ denotes the vector potential of the light field, with corresponding magnetic field $\vec{B} = \vec{\nabla} \times \vec{A}$ in Gaussian units.
The latter couples to the electronic spin magnetic moment which involves the vector of Pauli matrices $\vec{\sigma}=(\sigma_x,\sigma_y,\sigma_z)$.

In the scenario under consideration, the vector potential is given by
\begin{equation}
 \vec{A}(z,t) = f(t) \left[ \vec{A}_1(z,t) + \vec{A}_2(z,t) \right]
\label{eqn:vector-potential}
\end{equation}
with a right-travelling component
\begin{equation}
\vec{A}_1(z,t) = a_1 \vec{\varepsilon}_{1} \cos(\omega t - kz)
\end{equation}
and a left-travelling component of doubled frequency
\begin{equation}
\vec{A}_2(z,t) = a_2 \vec{\varepsilon}_{2} \cos(2\omega t + 2kz)\ .
\end{equation}
The first wave is characterized by the amplitude $a_1$ and the fundamental frequency $\omega = kc$. Its wave vector is given by $\vec{k} = k \vec{e}_z$.
The second wave has amplitude $a_2$ and oscillates with the second harmonic frequency.
Both fields are assumed to be linearly polarized along the $x$ axis, so that the polarization vectors are given by $\vec{\varepsilon}_{1}=\vec{\varepsilon}_{2}=\vec{e}_x$.
An envelope function $f(t)$ enters the total field in Eq.~\eqref{eqn:vector-potential} which allows switching on and off the field in the numerical calculations.
The overall laser intensity, when $f(t) \equiv 1$, is $I=\frac{\omega^2 \left( a_1^2 + 4 a_2^2 \right)}{8 \pi c}$ in this configuration.

We choose the incident electron momentum to be in the $y$--$z$ plane, so $\vec{p}\cdot\vec{A} = 0$.
As mentioned before, this geometry allows to highlight spin effects.
Having only $z$-dependence in the potential, Pauli's equation becomes effectively one-dimensional in space.
It can, thus, be solved by an ansatz in the form of an expansion into plane waves, or momentum eigenstates respectively:
\begin{eqnarray}
  \psi(t,z) = \sum_n c_n(t) e^{i n k z + i p_z z} = \sum_n c_n(t) \ket{n} \, .
\label{eqn:ansatz}
\end{eqnarray}
The electron spin is encoded in the time-dependent spinor expansion coefficients
$c_n = \left( \begin{matrix} c_n^\uparrow \\ c_n^\downarrow \end{matrix} \right)$.
The sum being discrete because, due to the periodicity of the potential, only the given discrete subset of momentum eigenstates do interact.
Here $p_z$ denotes the offset in the initial longitudinal electron momentum from an integer multiple of $k$.
By plugging Eq.~\eqref{eqn:ansatz} into Eq.~\eqref{eqn:tdpe} we obtain a coupled system of explicitly time-dependent ordinary differential equations:
\begin{eqnarray}
 i \dot{c}_n(t) = E_n c_n(t) + \mathcal{V}_n(t) + \mathcal{W}_n(t)
 \label{eqn:master_ode}
\end{eqnarray}
with the kinetic energies $E_n = \frac{(n k + p_z)^2}{2m}$ and
\begin{subequations}
\label{eqn:master_ode_parts}
\begin{eqnarray}
  \mathcal{V}_n(t) &=& \frac{e^2}{8 m c^2} f(t)^2 \bigg[
   a_2^2 e^{4 i \omega t} c_{n-4}(t)
   + 2 a_1 a_2 e^{i \omega t} c_{n-3}(t) \nonumber\\*
   &+& a_1^2 e^{-2 i \omega t} c_{n-2}(t)
   + 2 a_1 a_2 e^{3 i \omega t} c_{n-1}(t) \nonumber\\*
   &+& \left( 2 a_1^2 + a_2^2 \right) c_n(t) \nonumber\\*
   &+& 2 a_1 a_2 e^{-3 i \omega t} c_{n+1}(t)
   + a_1^2 e^{2 i \omega t} c_{n+2}(t) \nonumber\\*
   &+& 2 a_1 a_2 e^{-i \omega t} c_{n+3}(t)
   + a_2^2 e^{-4 i \omega t} c_{n+4}(t)
  \bigg] \label{eqn:master_ode_V}\\*
  \mathcal{W}_n(t) &=& \frac{i e \omega}{4 m c^2} \sigma_y f(t) \bigg[
   2 a_2 e^{2 i \omega t} c_{n-2}(t)
   + a_1 e^{-i \omega t} c_{n-1}(t) \nonumber\\*
   &-& a_1 e^{i \omega t} c_{n+1}(t)
   - 2 a_2 e^{-2 i \omega t} c_{n+2}(t)
  \bigg] \label{eqn:master_ode_W}
\end{eqnarray}
\end{subequations}
to encode the coupling to the neighboring states.
\subsection{Perturbative expansion by Dyson series}
\label{sub:dyson}
Based on the Pauli equation \eqref{eqn:tdpe} we can perform a perturbative treatment of the spin-dependent three-photon Kapitza-Dirac process, valid at small field amplitudes.
To this end, we consider energy-conserving transitions from momentum mode $\ket{-2}$ to momentum mode $\ket{2}$ assuming a vanishing initial momentum offset ($p_z = 0$).
From here on we use the abbreviation
\begin{equation}
 T := \int_{0}^{t_f} f(t) dt
\label{eqn:eff_T}
\end{equation}
to denote  an effective interaction time including switching on and off the fields at $t=0$ and $t=t_f$ respectively.
Specifically in this analytical consideration, the envelope function is set to $f(t) \equiv 1$.
In the Dyson series we need the free propagator
\begin{equation}
 U_0(t - t') = \sum_n e^{-i E_n \left( t - t' \right)} \ket{n}\bra{n}
\label{eqn:propagator}
\end{equation}
and the relevant terms in the potentials $V(t)=\frac{e^2}{2 m c^2} \vec{A}^2$ and $W(t)=\mu_{\rm B} \vec{\sigma} \cdot \vec{B}$ that can produce products proportional to $e^{4 i k z} = \sum_n \ket{n}\bra{n-4}$ with no time dependence [compare Eqs. \eqref{eqn:master_ode_V} and \eqref{eqn:master_ode_W}]:
\begin{subequations}
\label{eqn:V_relevant}
\begin{eqnarray}
 V_1(t) &=& \frac{e^2}{8m c^2} a_1^2 e^{-2 i \omega t} \sum_n \ket{n}\bra{n-2} \label{eqn:V_1} \\*
 V_2(t) &=& \frac{e^2}{4m c^2} a_1 a_2 e^{i \omega t} \sum_n \ket{n}\bra{n-3} \label{eqn:V_2} \\*
 W_1(t) &=& \frac{i e \omega}{4 m c^2} a_1 \sigma_y e^{-i \omega t} \sum_n \ket{n}\bra{n-1} \label{eqn:W_1} \\*
 W_2(t) &=& \frac{i e \omega}{2 m c^2} a_2 \sigma_y e^{2 i \omega t} \sum_n \ket{n}\bra{n-2} \label{eqn:W_2}
\end{eqnarray}
\end{subequations}
Expanding the transition amplitude from $\ket{-2}$ to $\ket{2}$ in a Dyson series up to third order in the amplitudes $a_{1,2}$ gives
\begin{widetext}
\begin{eqnarray}
 &&\bra{2}U(T)\ket{-2} \nonumber\\*
  &\approx& - \int_0^T dt_1 \int_0^{t_1} dt_2 \bra{2} U_0(T - t_1) [V(t_1) + W(t_1)] U_0(t_1 - t_2) [V(t_2) + W(t_2)] U_0(t_2) \ket{-2} \nonumber\\*
  &+& i \int_0^T dt_1 \int_0^{t_1} dt_2 \int_0^{t_2} dt_3 \bra{2} U_0(T - t_1) [V(t_1) + W(t_1)] U_0(t_1 - t_2) [V(t_2) + W(t_2)] U_0(t_2 - t_3) [V(t_3) + W(t_3)] U_0(t_3) \ket{-2} \nonumber\\*
  &\approx& - \frac{ie^3 \omega}{16 m^2 c^4} a_1^2 a_2 \sigma_y \left( I_+ + I_- + J_+ + J_- \right) + \frac{e^3 \omega^3}{32 m^3 c^6} a_1^2 a_2 \sigma_y \left( K_1 + K_2 + K_3 \right)
\end{eqnarray}
\end{widetext}
Note that the matrix element $\bra{2}U(T)\ket{-2}$ represents only the partial trace over the spatial dependence.
Thus the result remains in the algebra of Pauli matrices thereby encoding the full spin dynamics.
Each of the $I_\pm,J_\pm,K_{1,2,3}$ stands for one allowed combination of interaction terms in $V(t) + W(t)$.
For example by first applying $W_2$ at $t_2$ and then $V_1$ at $t_1$ or vice versa [see Eqs.~\eqref{eqn:W_2} and \eqref{eqn:V_1}] in the second-order perturbation integral we obtain
\begin{eqnarray}
 I_\pm &=& \int_0^T dt_1 \int_0^{t_1} dt_2 e^{-i E_2 T} e^{-i\left( -E_2 \pm 2 \omega \right) t_1} e^{-i \left( E_2 \mp 2\omega \right) t_2} \nonumber\\*
 &\approx& \frac{iT e^{-i E_2 T}}{E_2 \mp 2\omega}
\end{eqnarray}
where we have neglected all terms not linearly growing in $T$.
In the same manner we find
\begin{eqnarray}
 J_\pm &=& \int_0^T dt_1 \int_0^{t_1} dt_2 e^{-i E_2 T} e^{-i\left( -E_2 + E_1 \pm \omega \right) t_1} e^{-i \left( E_2 - E_1 \mp \omega \right) t_2} \nonumber\\*
 &\approx& \frac{iT e^{-i E_2 T}}{E_2 - E_1 \mp \omega}
\end{eqnarray}
for combinations of $V_2$ and $W_1$ [see Eqs.~\eqref{eqn:V_2} and \eqref{eqn:W_1}].
Finally,
\begin{subequations}
\begin{eqnarray}
 K_1 &=& \int_0^T dt_1 \int_0^{t_1} dt_2 \int_0^{t_2} dt_3 \nonumber\\*
 &&e^{-i E_2 T} e^{-i (-E_2 - 2 \omega) t_1} e^{-i (E_1 + \omega) t_2} e^{-i (E_2 - E_1 + \omega) t_1} \nonumber\\*
 &\approx& \frac{-T e^{-i E_2 T}}{\left( E_2 - E_1 + \omega \right)\left( E_2 + 2 \omega \right)} \\*
 K_2 &\approx& \frac{-T e^{-i E_2 T}}{\left( E_2 - E_1 + \omega \right)\left( E_2 - E_1 - \omega \right)} \\*
 K_3 &\approx& \frac{-T e^{-i E_2 T}}{\left( E_2 - 2 \omega \right)\left( E_2 - E_1 - \omega \right)}
\end{eqnarray}
\end{subequations}
combining twice $W_1$ \eqref{eqn:W_1} and once $W_2$ \eqref{eqn:W_2} in the third order of the Dyson series.
Thus we arrive at
\begin{eqnarray}
 &&\bra{2}U(T)\ket{-2} \nonumber\\*
 &\approx& \frac{e^3 \omega}{16 m^2 c^4} a_1^2 a_2 \sigma_y T e^{-i E_2 T} \left[ \frac{m c^2}{\omega^2 - m^2 c^4} + \frac{3 m c^2}{\frac{9}{4}\omega^2 - m^2 c^4} \right] \nonumber\\*
 &-&\frac{e^3 \omega^3}{32 m^3 c^6} a_1^2 a_2 \sigma_y T e^{-i E_2 T} \frac{5m^2 c^4}{2\left( \frac{9}{4} \omega^2 - m^2 c^4 \right)\left( \omega^2 - m^2 c^4 \right)} \nonumber\\*
 &=& \frac{e^3 \omega}{4m^3 c^6} a_1^{2} a_2 \sigma_y T e^{-i E_2 T} \frac{m^2 c^4}{\frac{9}{4}\omega^2 - m^2 c^4} \nonumber\\*
 &\approx& -\frac{e^3 \omega}{4m^3 c^6} a_1^2 a_2 \sigma_y T e^{-i E_2 T}
\end{eqnarray}
From there we see that only the spin flipping transition is allowed and we can deduce its Rabi frequency
\begin{equation}
 \Omega_R = \frac{e^3 \omega}{2m^3 c^6} a_1^2 a_2
\label{eqn:Rabi-freq}
\end{equation}
determining the short-time behavior of the spin-dependent scattering probability.
As we shall see below, in certain parameter regimes the latter adopts the form
\begin{equation}
 | c_{2}^\downarrow (T) |^2 = \sin^2 \left( \frac{1}{2} \Omega_R T \right)
 \label{eqn:rabi_cycle}
\end{equation}
if we start from $c_{-2}^\uparrow(0) = 1$.
\subsection{Effective ponderomotive potential}
\label{sub:effective_potential}
In this section we identify an effective ponderomotive potential arising from the vector potential \eqref{eqn:vector-potential}.
Instead of using a Dyson series, we exactly express the time evolution operator for Pauli's equation \eqref{eqn:tdpe} as a time ordered exponential
\begin{equation}
 U(T) = \Texp\left[ -i \int_{0}^{T} H(t) dt \right] =: \exp\left[ -i \mathcal{M}(T) \right] \, .
\end{equation}
In the second step, the Magnus expansion \cite{Magnus1954} has been applied,
where $\mathcal{M}(T) = \sum_i \mathcal{M}_i(T)$ is split into orders of powers of the Pauli Hamiltonian $H$.
Each of the $\mathcal{M}_i(T)$ is Hermitian of its own.
The same approach has been followed in \cite[Sec.~III]{Erhard2015} for 2-photon Kapitza-Dirac scattering up to terms in $\mathcal{M}_2$.
Relativistic corrections to Pauli's equation were required there, because elliptically polarized light was considered.
While in the present case of linear field polarization this is not necessary (see also App.~\ref{sec:volkov}),
we need to include the third order of the Magnus expansion to describe 3-photon interactions.
These orders are given by
\begin{subequations}
\begin{eqnarray}
 \mathcal{M}_1(T) &=& \int_{0}^{T} dt_1 H(t_1) \\*
 \mathcal{M}_2(T) &=& - \frac{i}{2} \int_{0}^T dt_1 \int_{0}^{t_1} dt_2 \left[ H(t_1), H(t_2) \right] \\*
 \mathcal{M}_3(T) &=& - \frac{1}{6} \int_{0}^T dt_1 \int_{0}^{t_1} dt_2 \int_{0}^{t_2} dt_3 \nonumber\\*
 &&\left\{ \left[ H(t_1), \left[ H(t_2), H(t_3) \right] \right] + \left[ \left[ H(t_1), H(t_2) \right], H(t_3) \right] \right\} \, . \nonumber\\*
\end{eqnarray}
\end{subequations}
With the help of computer algebra we find, that in our specific setup, where
$H(t) = -\frac{1}{2m}\frac{\partial^2}{\partial z^2} + V(t) + W(t)$ ($f(t)\equiv 1$), and $\vec{p}$ lying in the $y$--$z$-plane,
$\mathcal{M}_1(T) + \mathcal{M}_2(T) + \mathcal{M}_3(T)$ is asymptotically equal to $H_{\mathrm{eff}} T$
with the time-independent effective Hamiltonian
\begin{equation}
 H_{\mathrm{eff}} = - \frac{1}{2m} \frac{\partial^2}{\partial z^2} + \frac{e^3 \omega}{2 m^3 c^6} a_1^2 a_2 \sigma_y \sin(4 k z) \, .
 \label{eqn:H_eff}
\end{equation}
The latter consists of the well-known kinetic energy term and an effective ponderomotive potential.
Note that we neglected spatially constant ponderomotive terms that can be removed by a gauge transformation.
The effective ponderomotive potential accounts for the spin flipping transition from $\ket{-2}$ to $\ket{2}$ (or vice versa)
with the same Rabi frequency $\Omega_R = \frac{e^3 \omega}{2 m^3 c^6}a_1^2 a_2$ as before [see Eq.~\eqref{eqn:Rabi-freq}].

We note that the quantum dynamics of the two-state system satisfying the Bragg condition, which arises from this effective Hamiltonian, is governed by the system of coupled differential equations
\begin{subequations}
\begin{eqnarray}
 i \dot{c}_{-2}(t) &=& E_2 c_{-2}(t) + \frac{i}{2} \Omega_R \sigma_y c_{2}(t) \\
 i \dot{c}_{2}(t) &=& E_2 c_{2}(t) - \frac{i}{2} \Omega_R \sigma_y c_{-2}(t)
\end{eqnarray}
\end{subequations}
It resembles the two-state dynamics of the usual Kapitza-Dirac effect in the Bragg regime (see Eq.~(8) in \cite{Batelaan2007}) and,
in the present case, gives rise to the spin flipping Rabi oscillation mentioned in Eq.~\eqref{eqn:rabi_cycle}.

\section{Numerical results}
\label{sec:numerical}
In this section, we present our numerical results on spin-dependent Kapitza-Dirac scattering in bichromatic counterpropagating laser fields,
as described by the vector potential \eqref{eqn:vector-potential}.
Our main goal is to discuss the time evolution of the scattering probabilities.
This way, we extend the results presented in \cite{Mcgregor2015},
where total scattering probabilities have been obtained in counterpropagating bichromatic laser pulses with Gaussian profiles.
While the perturbative consideration in Sec.~\ref{sub:dyson} [see also Sec.~\ref{sub:effective_potential} and App.~\ref{sec:volkov}] have already revealed the short-time behavior of the scattering probabilities,
our numerical analysis will provide a comprehensive picture of the electronic time evolution.

We have used two different methods to solve the time-dependent Pauli equation \eqref{eqn:tdpe} numerically.
On the one hand, we propagate the equation directly in coordinate space by Fourier split-step methods.
On the other hand, we solve the coupled system of ordinary differential equations in momentum space \eqref{eqn:master_ode} by Runge-Kutta algorithms.
In both cases, we use a flat-top switching function $f(t)$ with sine-squared edges.

\subsection{Scattering of an electron wave packet}
\label{sub:wave_packet}
\begin{figure}[h]
\begin{center}
\includegraphics{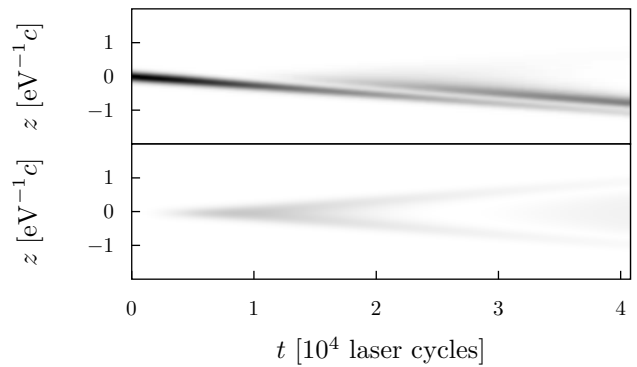}
\end{center}
\caption{Spin-dependent Kapitza-Dirac scattering of an electron wave packet with central momentum $-2k$ and width $0.1~\mathrm{eV}^{-1}c$.
 The spatial occupation probabilities of the upper (incident) and lower (scattered) spinor component of the electron wave function are shown in the corresponding panels.
 The field parameters are $\omega = 10^3~\mathrm{eV}$ and $e a_1 = e a_2 = 2 \times 10^4~\mathrm{eV}$.}
\label{fig:wave_packet}
\end{figure}

To begin with, we consider the one-dimensional dynamics of an electron wave packet in the bichromatic laser field.
The wave packet is assumed to be Gaussian shaped, having a central momentum of $-2 k$.
Its spin is prepared in the positive $z$-direction.
Its time evolution after entering the laser field is shown in Fig.~\ref{fig:wave_packet} in coordinate space.
Note that, in contrast to the interaction time $T$ [see Eq.~\eqref{eqn:eff_T}], small $t$ denotes times during a single interaction.
One can see the electron wave packet starting in a defined spin state at $t=0$.
After being partly scattered into the spin flipped state at $t \approx 10^4$ laser cycles it is being scattered back and forth between the two states.

\begin{figure}[h]
\begin{center}
\includegraphics{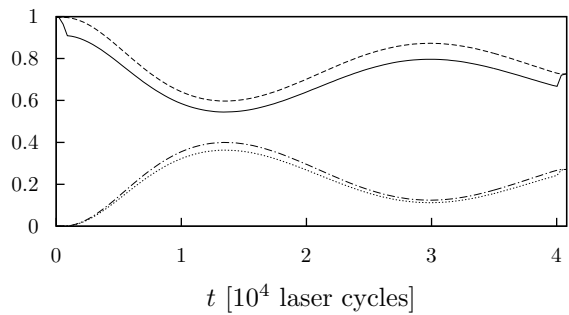}
\end{center}
\caption{
 Probabilities for the electron to be in the incident spin state and around the incident momentum $-2 k$ (solid line)
 as well as to be in the flipped spin state with the scattered momentum $2 k$ (dotted line).
 The interaction parameters are the same as in Fig.~\ref{fig:wave_packet}.
 Note that these probabilities add up to one only before switching on and after switching off the fields.
 Also shown are the total probabilities (summed over momenta) to be in the incident spin state (dashed line) and in the flipped spin state (dash-dotted line).
}
\label{fig:wave_packet_momentum}
\end{figure}

The corresponding time evolution in momentum space is shown in Fig.~\ref{fig:wave_packet_momentum}.
By comparing the solid and dashed lines, one can see, that first the spectrum of momenta is slightly broadened by switching on the electromagnetic field.
This can be understood as dressing by virtually absorbing and emitting photons.
Then a Rabi oscillation with simultaneous spin and momentum flip develops.
The amplitude of this oscillation is damped over time.
In the end, when switching off the fields, the dressing ceases and the electron is left in a quantum state where its momentum is entangled with its spin.

The dampening of the Rabi cycle can be readily understood by the fact,
that the wave packet with resonant central momentum involves also components with off-resonant momenta.
The latter contribute with slightly different Rabi frequencies.

\subsection{Rabi oscillation dynamics}
\label{sub:rabi_dynamics}
\begin{figure}[h]
\begin{center}
\includegraphics{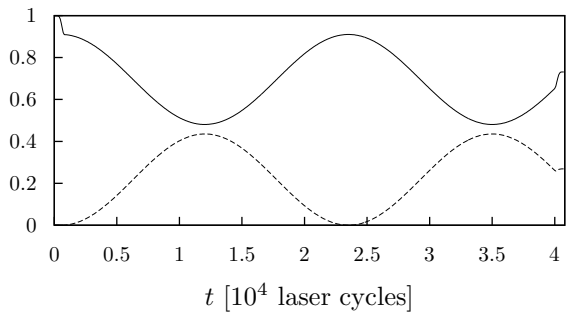}
\end{center}
\caption{Time evolution of the occupation probabilities $|c_{-2}^\uparrow(t)|^2$ (solid line) and $|c_{2}^\downarrow(t)|^2$ (dashed line)
for Kapitza-Dirac scattering from bichromatic counterpropagating waves with
$\omega = 10^{3} \,\mathrm{eV}$, $ e a_1 = e a_2 = 2 \times 10^{4}\,\mathrm{eV}$.
The combined laser intensity is $I = 6.82 \times 10^{21}\,\mathrm{W/cm^2}$.
The electron is incident with a longitudinal momentum of $-2 k$;
its transverse momentum is oriented perpendicularly to the field polarization and has an arbitrary magnitude.
}
\label{fig:time_evol}
\end{figure}

The oscillating population dynamics visible in Figs.~\ref{fig:wave_packet} and \ref{fig:wave_packet_momentum} can be highlighted more strongly
when scattering of plane-wave electron states with definite, resonant momentum is considered.
An example is shown in Fig.~\ref{fig:time_evol},
which illustrates the temporal evolution of the occupation probabilities $|c_{-2}^\uparrow(t)|^2$ and $|c_{2}^\downarrow(t)|^2$ during the interaction in the presence of the bichromatic laser field.
The electron starts with initial longitudinal momentum $-2 k$ and spin polarization along the $z$-direction.
While the effect of dressing in the field is still clearly pronounced, the damping of the Rabi cycle with passage of time is no longer visible.

\begin{figure}[h]
\begin{center}
\includegraphics{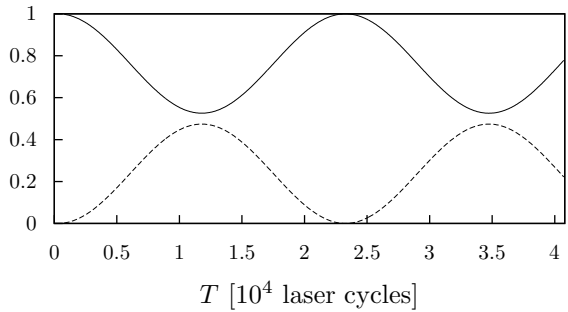}
\end{center}
\caption{Final occupation probabilities $|c_{-2}^\uparrow(T)|^2$ (solid line) and $|c_{2}^\downarrow(T)|^2$ (dashed line) for Kapitza-Dirac scattering
from bichromatic counterpropagating laser waves, as function of the interaction time $T$.
The electron and field parameters are chosen as in Fig.~\ref{fig:time_evol}.
Every plotted data corresponds to a full interaction with switching on and
off of $8 \times 10^{2}$ laser cycles each.}
\label{fig:time_rabi}
\end{figure}

An even cleaner picture of the Rabi oscillation dynamics is obtained when the dependence of the occupation probabilities
$|c_{-2}^\uparrow(T)|^2$ and $|c_{2}^\downarrow(T)|^2$ as a function of the interaction time $T$ is considered.
Our corresponding results are shown in Fig.~\ref{fig:time_rabi} where each plotted data point represents a full interaction including switching on and off the fields.
We note that the Rabi cycle is not fully developed,
and that the frequency is substantially higher than $\Omega_R = 3 \times 10^{-2}~\mathrm{eV} = 1.9 \times 10^{-4}~[\mathrm{laser\ cycles}]^{-1}$ as predicted by \eqref{eqn:Rabi-freq}.
The laser cycles always refer to the fundamental frequency $\omega$.

\begin{figure}[h]
\begin{center}
\includegraphics{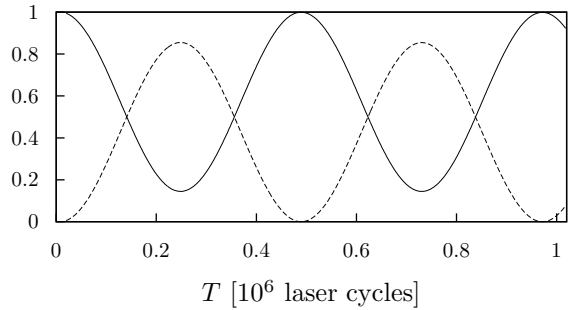}
\end{center}
\caption{Same as Fig.~\ref{fig:time_rabi} but with
$\omega = 2 \times 10^{2} \,\mathrm{eV}$ and $e a_1 = e a_2 = 8 \times 10^{3}\,\mathrm{eV}$, resulting in a laser intensity of $I = 4.36 \times 10^{19}\,\mathrm{W/cm^2}$.
Note that the Rabi cycles are more developed, here with $C \approx 0.855$.}
\label{fig:time_rabi_reduced}
\end{figure}

\begin{figure}[h]
\centering
\includegraphics{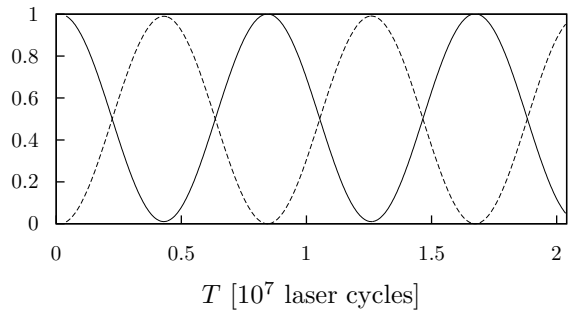}
\caption{
 Same as Fig.~\ref{fig:time_rabi} but with $\omega = 2.5 \times 10^{2}\,\mathrm{eV}$, $e a_1 = 4 \times 10^{3}\,\mathrm{eV}$ and $e a_2 = 2 \times 10^{3}\,\mathrm{eV}$,
 resulting in a laser intensity of $I = 6.82 \times 10^{18}\,\mathrm{W/cm^2}$.
 Here, the Rabi cycles are almost fully developed ($C \approx 1$).
}
\label{fig:time_rabi_asym}
\end{figure}

Nevertheless, to a very good approximation, the scattering probability can be described by
\begin{equation}
 | c_{2}^{\downarrow} (T) |^{2} = C \sin^2 \left( \frac{1}{2} \Omega T \right)
\label{eqn:rabi_reduced_cycle}
\end{equation}
instead of \eqref{eqn:rabi_cycle}, where $C$ is the maximally reached scattering probability.
Here, an effective Rabi frequency
\begin{equation}
\Omega
= \frac{1}{\sqrt{C}} \Omega_R
\label{eqn:eff_rabi}
\end{equation}
which accounts for the faster oscillation dynamics \cite{Awwad2016} has been introduced.
This reestablishes the agreement with \eqref{eqn:Rabi-freq} because,
for small times $T \ll \Omega^{-1}$, the factor $C$ drops out and we obtain $| c_2^{\downarrow} (T) |^2 \approx (\frac{1}{2} \Omega_R T)^2$ [see also App.~\ref{sec:reduced_rabi}].
According to our numerical calculations, this behavior, that the maximal scattering probability remains less than one, is rather typical.
In Fig.~\ref{fig:time_rabi} we have $C \approx 0.474$. \\
Further examples with different values of $C$ are shown in Figs.~\ref{fig:time_rabi_reduced} and \ref{fig:time_rabi_asym}.
The former is calculated in the same way as Fig.~\ref{fig:time_rabi}, but with longer laser wave length and lower amplitude, leading to more pronounced Rabi oscillations.
The latter is an example with different amplitudes of the counterpropagating waves.
It shows, that we can find parameters, where $C$ gets arbitrarily close to one.

\begin{figure}[h]
\begin{center}
\includegraphics{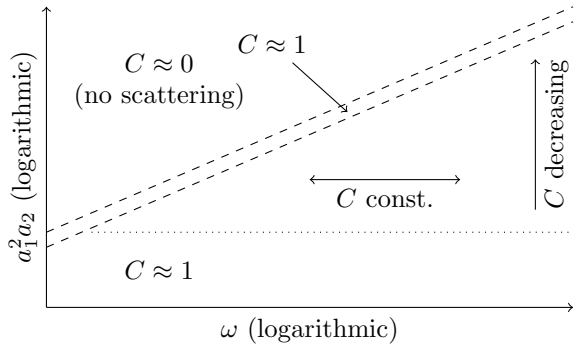}
\end{center}
\caption{
 Schematic diagram of the maximal Rabi amplitude $C$ in the $a_1^2 a_2$--$\omega$ plane indicating
 various parameter regimes.
 The position of the dyke summit is like $e^3 a_1^2 a_2 \approx \omega^2 10^{6}~\mathrm{eV}$ (order of magnitude).
}
\label{fig:scheme}
\end{figure}

We have performed a parameter scan in order to reveal for which values of field amplitude and frequency the Rabi oscillations are fully developed.
Our results are summarized as a schematic diagram in Fig.~\ref{fig:scheme}.
It shows that Rabi oscillations with maximum amplitude $C \approx 1$ are found in the bottom region of small laser amplitudes.
Increasing the laser amplitudes from there lowers $C$ substantially but independent of $\omega$ until the lower dashed line is reached.
Crossing that line, $C$ rises to $1$ again and then rapidly decreases to zero forming a dyke like structure.
Beyond the dyke practically no scattering takes place.
We found that \eqref{eqn:eff_rabi} holds for all parameter sets below and on top of the dyke from $\omega = 10~\mathrm{eV}$ to $3 \times 10^{3}~\mathrm{eV}$ and
from $e a_{1,2} = 2 \times 10^{3}~\mathrm{eV}$ to $4 \times 10^{4}~\mathrm{eV}$.

The reduced Rabi amplitudes may be caused by the fact that an intrinsic detuning develops in the system when the field amplitudes increase
(a corresponding analytical model is provided in appendix \ref{sec:reduced_rabi}, see also \cite{Ahrens2013}).
This line of argument would explain why $C\approx 1$ is reached for small field amplitudes where dressing effects are negligibly small.
In general, the dressings experienced by the incoming and outgoing electron, respectively, may slightly differ from each other since the field configuration is asymmetric.
As a consequence, small mismatches in the energy-momentum balance can arise when the values of $a_{1,2}$ increase,
leading to damped Rabi oscillations.
However, for very specific values of $a_1$, $a_2$ and $\omega$ the dressing effects could be sizeable but coincide for the incoming and outgoing electron.
This circumstance might explain the appearance of the dyke-like structure where $C \approx 1$ is reached again.
The position of the dyke summit is approximately described by $e^3 a_1^2 a_2 \approx \omega^2 10^{6}~\mathrm{eV}$.

Let us put our results into an experimental context.
To be specific, we consider the example shown in Fig.~\ref{fig:time_rabi}.
The corresponding laser field frequencies ($\omega \sim \mathrm{keV}$) and intensities ($I \sim 10^{21}~\mathrm{W/cm^2}$) can,
in principle, be attained from high-power x-ray free-electron lasers such as the European XFEL (Hamburg, Germany) or LCLS (Stanford, California).
The latter currently reaches x-ray intensities up to $10^{20}~\mathrm{W/cm^2}$ at $\sim 100~\mathrm{fs}$ pulse durations and $\sim 1~\mu\mathrm{m}$ focal widths \cite{Bostedt2016}.
For the present purpose it is important that the spatiotemporal extension of the laser fields is sufficiently large.
For instance, the most pronounced effect in Fig.~\ref{fig:time_rabi} occurs after an interaction time of about $T \approx 50\mathrm{fs}$,
which corresponds to a focal width of $\Delta y = vT$ to be traversed by the electron.
Here, $v$ denotes the transverse electron velocity.
For electron energies in the $\mathrm{eV}$ range, we obtain $\Delta y \sim 1 \mu\mathrm{m}$.
The required laser pulse energy, accordingly, can be estimated by order of magnitude as
\begin{equation}
\mathcal{E} \approx I \Delta x \Delta y \Delta \tau
\sim \frac{(mc^2)^5}{\alpha e^4 a_1^2} \left( \frac{4}{a_1^2} +
\frac{1}{a_2^2} \right) \frac{\Delta x}{\lambda_e} \, ,
\end{equation}
with the pulse duration $\Delta \tau \gtrsim T$, finestructure constant $\alpha \approx \frac{1}{137}$, and electron de Broglie wave length $\lambda_e$.

In order to observe the spin flips, a polarized incident electron beam can be used.
The spin state of the outgoing electrons can be probed by integrating, e.g., a Mott spectrometer into the detection system.
By varying the interaction time with the laser fields from run to run, the spin oscillation shown in Fig.~\ref{fig:time_rabi} could be resolved.

\section{Spin-dependent Kapitza-Dirac scattering of protons}
\label{sec:proton}
Apart from electrons, also other particles can undergo Kapitza-Dirac scattering.
As an example, we shall consider Kapitza-Dirac scattering of protons in this section.
Being spin-$\frac{1}{2}$-particles like electrons, the scattering dynamics of protons may be spin dependent, as well.
Our goal is to draw a comparison between (spin-dependent) Kapitza-Dirac scattering of protons versus electrons.
Both particles differ by their mass, charge, and $g$-factor.

Let $\Psi_\mathrm{p} = \Psi_\mathrm{p}(\vec{x},t)$ denote the two-component spinor wave function of the proton.
In the nonrelativistic regime, its time evolution is governed by the Pauli equation
\begin{eqnarray}
 i \frac{\partial}{\partial t} \Psi_\mathrm{p}
 = \left[ \frac{1}{2 m_\mathrm{p}} \left( -i \vec{\nabla} - \frac{e}{c} \vec{A} \right)^2
 + \frac{1}{2} g_\mathrm{p} \mu_\mathrm{N} \vec{\sigma} \cdot \vec{B} \right] \Psi_\mathrm{p}\,,
\label{eqn:tdpe-protons}
\end{eqnarray}
with the proton mass $m_\mathrm{p}$,
proton $g$-factor $g_\mathrm{p}\approx 5.5857$ and the nuclear magneton $\mu_\mathrm{N} = \frac{e}{2m_\mathrm{p}c}$.
The vector potential $\vec{A}$ and magnetic field $\vec{B}$ in Eq.~\eqref{eqn:tdpe-protons} are assumed to involve a wave vector $\vec{k}$.

Equation~\eqref{eqn:tdpe-protons} has the same structure as Eq.~\eqref{eqn:tdpe} for the electron.
Their mutual relation can be made explicit by a scaling transformation of the coordinates, according to \begin{eqnarray}
\vec{x}_\mathrm{e} = \frac{\vec{x}}{\rho}\,,\ \ t_\mathrm{e} = \frac{t}{\rho}\,,
\label{eqn:scaling}
\end{eqnarray}
with the electron-to-proton mass ratio $\rho = m_\mathrm{e}/m_\mathrm{p}$.
When rewritten in the scaled coordinates, the Pauli equation \eqref{eqn:tdpe-protons} becomes
\begin{eqnarray}
 i \frac{\partial}{\partial t_\mathrm{e}} \Psi
 = \left[\frac{1}{2 m_\mathrm{e}} \left( -i \vec{\nabla}_\mathrm{e} + \frac{e}{c} \vec{A}_\mathrm{e} \right)^2
 - \frac{1}{2} g_\mathrm{p} \mu_\mathrm{B} \vec{\sigma} \cdot \vec{B}_\mathrm{e} \right] \Psi \,,
\label{eqn:tdpe-scaled}
\end{eqnarray}
with the wave function $\Psi = \Psi(\vec{x}_\mathrm{e},t_\mathrm{e})$ now depending on the scaled space and time coordinates $\vec{x}_\mathrm{e}$ and $t_\mathrm{e}$.
Besides, $m_\mathrm{e}$ is the electron mass,
and $\mu_\mathrm{B} = \frac{e}{2 m_\mathrm{e} c}$ the Bohr magneton.
The electromagnetic field parameters after scaling read
\begin{eqnarray}
 \vec{A}_\mathrm{e} = -\rho \vec{A} \,,
 \vec{B}_\mathrm{e} = \vec{\nabla}_\mathrm{e} \times \vec{A}_\mathrm{e} = -\rho^2 \vec{B} \,,
 \vec{k}_\mathrm{e} = \rho \vec{k} \,.
\end{eqnarray}
Equation~\eqref{eqn:tdpe-scaled} is almost identical to the Pauli equation which describes the time evolution of an electron in the scaled electromagnetic field.
In fact, if we ignored the spin interaction term, the corresponding Schr\"odinger equations would coincide.
Consequently, the Kapitza-Dirac scattering dynamics of electrons and protons is fully equivalent to each other in situations when the particle spin is immaterial.
For instance, the experimental verification of the Kapitza-Dirac effect on electrons applied optical laser fields
of frequency $\omega_\mathrm{e} \approx 2.3\,\mathrm{eV}$ and intensity $I_\mathrm{e} \sim 10^{9} \, \mathrm{W/cm^2}$.
An equivalent setup with protons could be realized with x-ray laser fields of frequency $\omega \approx 4.3 \, \mathrm{keV}$ and intensity $I \sim 10^{22} \, \mathrm{W/cm^2}$.

However, due to the spin interaction term, there exists one important difference between Eqs.~\eqref{eqn:tdpe-scaled} and \eqref{eqn:tdpe}.
The former still contains the proton $g$-factor $g_\mathrm{p}$, which differs from the electron $g$-factor $g_\mathrm{e}=-2$ by almost a factor of three in magnitude.
As a consequence, in spin-sensitive interaction processes, electron and proton dynamics are not fully equivalent.
For the particular case of spin-dependent Kapitza-Dirac scattering involving three-photons (see Sec.~\ref{sec:theory}),
the proton interacts more strongly with the -- correspondingly scaled -- electromagnetic fields than the electron.
The enhanced interaction strength can be quantified by the Rabi frequency which amounts to
\begin{eqnarray}
 \Omega_\mathrm{p} = g_\mathrm{p} \frac{e^3 a_1^2 a_2 \omega}{4 m_\mathrm{p}^3 c^6}
\end{eqnarray}
for protons.

\section{Conclusion}
\label{sec:conclusion}
Spin-dependent Kapitza-Dirac scattering of electrons and protons from bichromatic laser waves was studied in the nonrelativistic regime.
Our consideration focused on the case of linearly polarized counterpropagating laser waves with a fundamental frequency and its second harmonic.
On the one hand, we derived analytical results for the associated short-time scattering probability in this field configuration.
It was shown that the deflection of electrons with specific incident momenta is necessarily accompanied by a flip of the electronic spin components orthogonal to the magnetic field direction.
This way, a pronounced entanglement between the outgoing electron momentum and spin state arises.

On the other hand, the full time dependence of the electron dynamics was obtained by solving the time-dependent Pauli equation numerically.
The coherent scattering of the electrons leads to characteristic Rabi cycles.
We emphasize that, for equal amplitudes of the counterpropagating laser waves,
the same kind of Rabi oscillations were also found in \cite{Ahrens2012,Ahrens2013} in a strongly Doppler-shifted reference frame.
By performing a systematic parameter scan, covering a broad range from ultraviolet to X-ray laser frequencies,
we found and characterized different interaction regimes in which the Rabi cycles are either fully developed,
only partially developed or completely suppressed.

Due to numerical feasibility, we applied laser fields with rather high frequencies and intensities in our computations.
Radiation sources with corresponding characteristics are, in principle,
available through high-harmonic emission from laser-irradiated plasma surfaces \cite{Roedel2012,*Cerchez2013} or at free-electron laser laboratories,
such as the FLASH facility at DESY (Hamburg, Germany) \cite{desy} and the LCLS at SLAC (Stanford, California) \cite{LCLS}.
In the examples shown, the development of a substantial fraction of a Rabi cycle requires interaction times in the range of several femtoseconds up to picoseconds.

In a forthcoming study we intend to investigate spin-dependent Kapitza-Dirac scattering in laser waves of circular polarization,
where the photons carry a definite helicity (see also \cite{Erhard2015,*[{see also }]Bauke2014b,*Bauke2014a,Freimund2003}).

\section*{Acknowledgement}
 Inspiring discussions with M.\,J.\,A. Jansen are gratefully acknowleged.
 We thank S. Ahrens and H. Bauke for carefully reading the manuscript.
 This study was supported by SFB TR18 of the German Research Foundation (DFG) under project No. B11.

\appendix
\section{A model for field-induced detuning}
\label{sec:reduced_rabi}
In Sec.~\ref{sub:rabi_dynamics} we saw that the Rabi oscillations between the incident electron state and the spin-flipped scattered state are not always fully developed.
The reason was attributed to an intrinsic, field-induced detuning in the system.
In this appendix, we provide a simplified model which is able to demonstrate this behavior.

\begin{figure}[h]
 \includegraphics{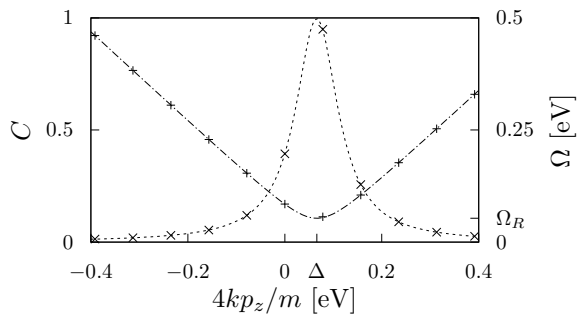}
\caption{Resonance behavior of the Rabi oscillation of the spin-flipping Kapitza-Dirac transition from state $\ket{-2 \uparrow}$ to state $\ket{2 \downarrow}$.
 The laser parameters read $\omega = 10^{3}~\mathrm{eV}$ and $e a_1 = e a_2 = 2.4 \times 10^{4} \mathrm{eV}$.
 The Rabi amplitude $C$ is shown (crosses) with a fit of Eq.~\eqref{eqn:rabi_amplitude} over $\Delta$ and $\Omega_R$ (dashed line).
 Additionally the measured frequency $\Omega$ (pluses) is compared to Eq.~\eqref{eqn:rabi_relation} (dash-dotted line) with the same fit parameters.
 The latter are marked on the corresponding axes.
\label{fig:resonance_peak}}
\end{figure}

Let us assume we can model our setup as a quantum two-state system in the orthonormal basis $\ket{-2 \uparrow}, \ket{2 \downarrow}$.
Additionally, we introduce a longitudinal momentum offset $p_z$ to both states giving a detuning in the kinetic energy of $\mp \frac{4 k p_z}{m}$.
The simplified Hamilton operator then reads
\begin{equation}
 H = \frac{1}{2} \left( \begin{matrix} \delta & \Omega_R \\ \Omega_R & -\delta \end{matrix} \right) \quad \mathrm{with} \quad \delta = \Delta - \frac{4 k p_z}{m} \, .
\label{eqn:hamilton}
\end{equation}
Here $\Delta$ describes an a-priori undetermined intrinsic detuning.
For the time evolution operator we obtain
\begin{eqnarray}
 &U(T) = \exp\left( -iTH \right) = \cos\left( \frac{\Omega}{2} T \right) - \frac{2 i}{\Omega} \sin\left( \frac{\Omega}{2} T \right) H \nonumber\\*
 &= \left( \begin{matrix}
  \cos\left( \frac{\Omega T}{2} \right) - \frac{i \delta}{\Omega} \sin\left( \frac{\Omega T}{2} \right) &
  - \frac{i \Omega_R}{\Omega} \sin\left( \frac{\Omega T}{2} \right) \\
  - \frac{i \Omega_R}{\Omega} \sin\left( \frac{\Omega T}{2} \right) &
  \cos\left( \frac{\Omega T}{2} \right) + \frac{i \delta}{\Omega} \sin\left( \frac{\Omega T}{2} \right)
 \end{matrix} \right) \nonumber\\*
\label{eqn:time_evolution}
\end{eqnarray}
with the effective Rabi frequency
\begin{equation}
 \Omega := \sqrt{\Omega_R^2 + \delta^2} = \sqrt{\Omega_R^2 + \left( \Delta - \frac{4 k p_z}{m} \right)^2} \, .
 \label{eqn:rabi_relation}
\end{equation}
The off-diagonal terms of \eqref{eqn:time_evolution} describe the transition amplitude from one state to the other.
Therefore we can infer, that the occupation of the states oscillates with the effective Rabi frequency $\Omega$ and amplitude
\begin{equation}
 C = \frac{\Omega_R^2}{\Omega^2} = \frac{1}{1 + \left( \frac{\Delta - 4 k p_z / m}{\Omega_R} \right)^2} \, .
 \label{eqn:rabi_amplitude}
\end{equation}
More explicitly, the transition probability is
\begin{equation}
 \left| \braket{2 \downarrow | U(T) | -2 \uparrow} \right|^2 = C \sin^2 \left( \frac{\Omega}{2} T \right) \, .
\label{eqn:transition}
\end{equation}
Thus, the Rabi oscillation amplitude is generally damped.
In Fig.~\ref{fig:resonance_peak} a sweep over the detuning parameter is shown for a parameter set at moderately high vector potential just below the dyke in Fig.~\ref{fig:scheme}.
It can be seen, that the numerical simulations based on Eq.~\eqref{eqn:master_ode} nicely resemble the detuning model described above,
and that the intrinsic detuning $\Delta$ is positive.
Note that very similar resonance peaks have been investigated in \cite{Ahrens2013} for monochromatic 2-photon and 3-photon Kapitza-Dirac scattering.
Further numerical simulations show, that $\Delta$ decreases with $k$, reaching the dyke summit when $\Delta = 0$ and drops quite fast beyond the dyke.
This is in good agreement with the structure described in Fig.~\ref{fig:scheme}.

Still, for short times, the transition probability is independent of the detuning and given by
$\left| \braket{2 \downarrow | U(T) | -2 \uparrow} \right|^2 = \frac{\Omega_R^2}{4} T^2 + o(\Omega_R^3 T^3)$.
This behaviour is also well-known from Rabi oscillations in two-level systems with permanent detuning \cite[Chap.~5.2.1]{Scully1997}.

Our model Hamiltonian \eqref{eqn:hamilton} thus catches the basic phenomenology of detuned Rabi oscillation dynamics.
We point out that it closely resembles the effective Hamiltonian in Eq.~\eqref{eqn:H_eff} which followed from the Magnus expansion.
As argued in Sec.~\ref{sec:numerical}, the physical origin of the field-induced detuning might be connected to the asymmetric field configuration,
with co- and counterpropagating laser waves of different frequencies (and amplitudes).
The relevant value of the detuning $\Delta$ in the numerical results of
Figs.~\ref{fig:time_rabi} and \ref{fig:time_rabi_reduced} can be inferred by comparing Eqs.~\eqref{eqn:eff_rabi} and \eqref{eqn:rabi_relation},
or by measuring the characteristics of the resonance peak like in Fig.~\ref{fig:resonance_peak}.

\section{Alternative approach based on Dirac-Volkov states}
 \label{sec:volkov}
In this appendix, we present an alternative derivation of the spin-flip Rabi frequency in Eq.~\eqref{eqn:Rabi-freq}.
Our approach is based on relativistic Volkov states which are solutions to the Dirac equation in the presence of a plane-wave laser field.
Usually, these states can only be applied to problems involving a single travelling laser wave.
In contrast, the spin-dependent Kapitza-Dirac effect under consideration occurs in the combined fields of two counterpropagating waves.
However, out of the three photons which are exchanged in total, only one is emitted into (or absorbed from) the high-frequency field mode.
In the limit of moderate laser intensities, the influence of this field mode can, thus, be treated within the first order of perturbation theory.
The interaction with the low-frequency field mode, in turn, may be incorporated into dressed electronic states.
While this kind of approach, in principle, allows to treat the impact of the low-frequency wave nonperturbatively,
we shall be interested in the perturbative limit where two low-frequency photons participate in the process.
In comparison with standard perturbation theory (of third order for both fields), this approach has the advantage of being rather compact.

The Kapitza-Dirac effect may be regarded as stimulated Compton scattering.
We therefore start our consideration from the usual $S$ matrix describing multiphoton Compton scattering (see, e.g.\@, \cite{Ivanov2004,Boca2012,Krajewska2013,*Krajewska2014})
\begin{equation}
 \mathcal{S} = \frac{ie}{c} \int d^4x\, \overline{\psi}_{p',s'} \slashed{A}_2 \psi_{p,s}\ .
\label{eqn:S}
\end{equation}
Here,
\begin{equation}
 \psi_{p,s}(x) = \sqrt{\frac{mc^2}{VE_p}} \left( 1 - \frac{e \slashed{k} \slashed{A}_1 (kx)}{2c (kp)} \right) u_{p,s}\,
 e^{-i(px)+i\Lambda_p}\ ,
\label{eqn:volkov}
\end{equation}
with
\begin{equation}
 \Lambda_p = \frac{1}{c(kp)} \int^{(kx)} \left[ e (p A_1(\phi)) + \frac{e^2}{2c} A_1^2(\phi) \right] d\phi
\end{equation}
denotes the Dirac-Volkov state for the incoming electron dressed by the field $A_1^\mu(\phi) = a_1 \cos(\phi)\,\varepsilon^\mu$.
Accordingly, $\psi_{p',s'}$ is the Dirac-Volkov state for the scattered electron.
We use the notation $(vw) = v^0 w^0 - \vec{v} \cdot \vec{w}$ for the product of two four-vectors $v^\mu = (v^0, \vec{v})$ and $w^\mu = (w^0, \vec{w})$.
Feynman slash notation is employed for the four-product with Dirac $\gamma$-matrices.
The free Dirac spinors $u_{p,s}$ are taken from \cite{Bjorken1964}, with the spin quantized along the $z$ axis.

For definiteness, we assume that the incident electron has a longitudinal momentum component $p_3= -2 k$ (zero offset) and a spin projection $s = \frac{1}{2}$.
In order to be scattered into the mirrored momentum state with $p_3' = 2 k$,
the electron needs to absorb two photons from the field $A_1^\mu$ and to emit one photon into the counterpropagating field $A_2^\mu$.
Besides, the electron momenta are assumed to be nonrelativistic, so that $E_p = E_{p'} \approx mc^2$ holds.

The integral in \eqref{eqn:S} can be evaluated by performing a Fourier series expansion, according to
\begin{eqnarray}
 e^{i(\Lambda_p-\Lambda_{p'})} &=& \exp\left\{ \frac{ie}{c} a_1\left[ \frac{(p\varepsilon)}{(kp)}-\frac{(p'\varepsilon)}{(kp')} \right]\sin(kx) \right.\nonumber \\
 & &\ \ \ - \left. \frac{ie^2}{8c^2} a_1^2\left[ \frac{1}{(kp)}-\frac{1}{(kp')} \right]\sin(2kx) \right\} \nonumber \\
 &=& \sum_n \tilde{J}_n(\alpha, \beta)\,e^{-in(kx)}\ .
\label{Fourier}
\end{eqnarray}
Here, the abbreviations
\begin{equation}
 \alpha = \frac{e}{c} a_1 \left[ \frac{(p' \varepsilon)}{(kp')} - \frac{(p \varepsilon)}{(kp)} \right] \ ,\ \
 \beta = \frac{e^2}{8c^2} a_1^2 \left[ \frac{1}{(k p)}-\frac{1}{(k p')} \right]
\end{equation}
have been introduced and the $\tilde{J}_n = \tilde{J}_n(\alpha, \beta)$ denote generalized Bessel functions \cite{[{}][{; Appendix B.}]Reiss1980}.
The sum in Eq.~\eqref{Fourier} runs over the number of photons exchanged with the laser field $A_1^\mu$.
The term with $n=1$ ($n=2$) corresponds to the absorption of one photon (two photons) from the wave $A_1^\mu$.

The relevant contribution to the $S$ matrix thus reads
\begin{eqnarray}
 \mathcal{S} &\approx& \frac{ie}{cV} \int d^4x\,\overline{u}_{p',s'}\bigg(\slashed{A}_2^{(+)} \tilde{J}_2\,e^{i (p' - p - 2k) x}
 - \frac{e}{2c}\slashed{A}_1^{(-)}\slashed{k}\slashed{A}_2^{(+)} \nonumber\\
 & & \times \left.\left[\frac{1}{(kp)} + \frac{1}{(kp')} \right] \tilde{J}_1\,e^{i(p' - p - k)x} \right) u_{p,s}\ .
\label{eqn:S2}
\end{eqnarray}
Here, $\slashed{A}_1^{(-)}=-\frac{1}{2} a_1 \gamma^1\,e^{-i(kx)}$ is understood as the part of $\slashed{A}_1$ which describes photon absorption.
Similarly, $\slashed{A}_2^{(+)}$ is the part of $\slashed{A}_2$ responsible for photon emission.
Note that $\slashed{A}_1^{(-)}\slashed{k}\slashed{A}_2^{(+)} = \slashed{A}_2^{(+)}\slashed{k}\slashed{A}_1^{(-)}$.
In the perturbative limit of small field amplitudes, the Bessel functions may be expanded \cite{[{}][{; Appendix B.}]Reiss1980}:
$\tilde{J}_1\approx \frac{\alpha}{2}$, $\tilde{J}_2\approx \frac{\alpha^2}{8} + \frac{\beta}{2}$.
Moreover, if we assume that the electron momentum has no component along the field polarization (i.e.\@, $\alpha=0$), Eq.~\eqref{eqn:S2} further simplifies to
\begin{equation}
\mathcal{S} \approx \frac{i e}{2 c V} \beta \int d^4x\,\overline{u}_{p',s'}\slashed{A}_2^{(+)} u_{p,s}\,e^{i(p'-p-2k)x}\ .
\label{eqn:S3}
\end{equation}
By an explicit evaluation one finds, that the spinor-matrix product vanishes identically if the spin quantum numbers coincide ($s=s'$).
In contrast, when the electron transition involves a spin flip ($s' = -\frac{1}{2}$), one obtains $\overline{u}_{p',s'}\gamma^1 u_{p,s}=- p_3/(mc)$.  
The space-time integration in Eq.~\eqref{eqn:S3} produces a factor $cVT$ since we have chosen the outgoing electron momentum $p'$ to agree with the energy-momentum conservation in the process.
Thus, the $S$ matrix involving a spin flip has the form
\begin{eqnarray}
\mathcal{S}_{\rm flip} \approx - \frac{i}{2} \Omega_R T\ ,
\end{eqnarray}
with
\begin{eqnarray}
\Omega_R = \frac{e}{2} \beta a_2\,\frac{p_3}{mc} \approx \frac{e^3 \omega}{2 m^3 c^6} a_1^2 a_2\ .
\end{eqnarray}
This result coincides with the Rabi frequency of Eq.~\eqref{eqn:Rabi-freq} giving a confirmation that the Pauli equation is sufficient
to treat the electron dynamics in the considered field configuration.
In contrast, in fields of circular polarization relativistic corrections to the Pauli Hamiltonian need to be taken into account \cite{Erhard2015,*Bauke2014b,*Bauke2014a}.

\end{document}